\documentclass[epj]{svjour}
\usepackage{graphics}
\usepackage{amsmath}
\usepackage{amssymb}
\usepackage{subcaption}
\usepackage{mathabx}
\usepackage{xspace}
\usepackage{hyperref}
\usepackage[titletoc, title]{appendix}
\usepackage[normalem]{ulem}

\begin{document}
\title{Inference of cosmic-ray source properties by conditional invertible neural networks}
\author{Teresa Bister\inst{1} \and Martin Erdmann\inst{1} \and Ullrich K\"othe\inst{2} \and Josina Schulte\inst{1}
}                     
\institute{RWTH Aachen University, III. Physikalisches Institut A, Otto-Blumenthal-Str., 52074 Aachen, Germany \and University of Heidelberg, Interdisciplinary Center for Scientific Computing,
Im Neuenheimer Feld 205, 69120 Heidelberg, Germany \and \email{erdmann@physik.rwth-aachen.de}}
\date{Received: date / Revised version: date}
%
\abstract{
The inference of physical parameters from measured distributions constitutes a core task in physics data analyses. Among recent deep learning methods, so-called \textit{conditional invertible neural networks} provide an elegant approach owing to their probability-preserving bijective mapping properties. They enable training the parameter-observation correspondence in one mapping direction and evaluating the parameter posterior distributions in the reverse direction. Here, we study the inference of cosmic-ray source properties from cosmic-ray observations on Earth using extensive astrophysical simulations. We compare the performance of conditional invertible neural networks (cINNs) with the frequently used Markov Chain Monte Carlo (MCMC) method. While cINNs are trained to directly predict the parameters' posterior distributions, the MCMC method extracts the posterior distributions through a likelihood function that matches simulations with observations. Overall, we find good agreement between the physics parameters derived by the two different methods. As a result of its computational efficiency, the cINN method allows for a swift assessment of inference quality.
} 
\maketitle

\section{Introduction}
\label{sec:introduction}

Information about our universe is obtained from four different messengers: electromagnetic waves, cosmic rays, neutrinos, and gravitational waves. Large observatories exist worldwide  for all these messengers. Each messenger carries information about its origin and, in addition, the history of its propagation to Earth as well as the distortion due to observational technology. To obtain dedicated information about properties of the universe, e.g., the source of the messenger, the influences of the other contributions (propagation, detector effects) have to be reversed by mathematical methods. This process is usually referred to as inference.

Among the four messengers, ultra-high-energy cosmic rays are characterized by two complicating effects in propagation and detection. First, the ionized nuclei can interact with cosmic background fields and may undergo directional deflections, lose energy, or decay sequentially on their way towards Earth. Second, as the nuclei enter the Earth's atmosphere, they cause showers of more than a billion secondary particles that allow only indirect determination of the primary's properties. Observatories enable an investigation of cosmic-ray frequency, the direction of arrival, energy, and cross-section, all of which can be used to unveil characteristics about the cosmic ray's origin.

In this work, we exemplarily investigate two methods for characterizing cosmic-ray sources from observations on Earth. Our focus is on correcting the measured energy distribution together with the distribution of the shower depths in the atmosphere for propagation effects. The measurements of today's observatories are so precise that bias and smearing of the detectors can be corrected comparatively easily \cite{AugerPRD2020,AugerCompositionICRC2019}. Instead of the complete true energy distribution at the source, we aim to determine a set of characteristic quantities at the source, describing the set of different atomic nuclei (composition), the power of the energy spectrum (spectral index), and the maximum accelerator energy. This astrophysical scenario is kept very simple, but already shows the sensitivity of the measurements to source properties of cosmic rays \cite{AugerCF}. This sensitivity is expressed in terms of posterior distributions of the characteristic source parameters.

Propagation of cosmic rays is associated with the above-mentioned interactions and nuclear decays, which follow each other sequentially in a random order. Propagation from source to observation is simulated by software such as CRPropa3~\cite{CRPropa3}. Because of the many random processes, the simulation can only be performed in a forward direction and inversion did not seem possible at first.

Therefore, the first determinations of the source properties were usually performed using forward simulations: the characteristic quantities of the sources (source parameters) were modified until the measured distributions on Earth were reproduced by the forward simulations after cosmic-ray propagation. To avoid simulating each parameter setting individually, databases for the astrophysical scenario with different parameter settings were created. These databases contain weight factors for each measured spectrum and allow interpolation between the simulated parameter settings. Finally, to efficiently search for associated source parameters and their posterior distributions with the measured distributions, one could use Bayesian methods like the Markov Chain Monte Carlo (MCMC).

With regard to new developments in the context of neural networks, the inversion of the above-described propagation of cosmic rays is now possible after all. So-called normalizing flow networks consist of invertible blocks which enable network training in the forward direction and an evaluation in the backward direction, while preserving probability in both directions \cite{NormalizingFlows,2019arXiv190809257K,papamakarios2021normalizing}. During the training process, a database of forward simulations in the one direction of computation is used, similarly as in Bayesian methods. The evaluation of a measured distribution by the trained network however happens in the backward computational direction, where the output consists of the posterior distributions of the source parameters. Recently, a number of inference methods based on deep learning methods have been developed and investigated (refer to the collection in \cite{Feickert:2021ajf}), including the unfolding of particle distributions \cite{Bellagente:2020piv} and the characterization of spatially correlated $\gamma$-ray maps \cite{Cranmer:2021}. 

In this paper, we present a normalizing flow network for the determination of cosmic-ray source parameters from measured distributions. The quality of this so-called \textit{conditional invertible neural network} (cINN) is investigated in a comparative study with the traditional MCMC method.

This work is structured as follows: First, we introduce the astrophysical scenario and describe how we generated the database for the mapping between source parameters and observed distributions. Then, we briefly recall the MCMC procedure before going into detail about the functionalities of the cINN. In the subsequent section we then compare the results of the two methods and the computing resources used. Finally, we present our conclusions.

\section{Astrophysical scenario and database}
\label{sec:astrophysical_model}
The astrophysical scenario used to investigate the performance of the inference methods is based on the findings of~\cite{AugerCF}. It consists of homogeneous sources which isotropically emit cosmic rays with different mass numbers $A_\mathrm{inj}$ and corresponding charge numbers $Z_\mathrm{inj}$. Acceleration is possible up to a maximum rigidity $R_\mathrm{cut}$ where rigidity denotes energy divided by charge: $R=E/Z$. The cosmic-ray emission with energy $E_\mathrm{inj}$ is described by a power law with spectral index $\gamma$ and a broken exponential cutoff:

\begin{equation}
\label{eq:emission}
\begin{split}
    J_\mathrm{inj}(E_\mathrm{inj}, A_\mathrm{inj}) = J_0 \cdot a(A_\mathrm{inj}) \Big(\frac{E_\mathrm{inj}}{10^{18}~\mathrm{eV}} \Big)^\mathrm{-\gamma} ~\\\\
    \cdot \begin{cases}
    1 & Z_\mathrm{inj} R_\mathrm{cut} < E_\mathrm{inj} \\
    \exp \big( 1-\frac{E_\mathrm{inj}}{Z_\mathrm{inj} R_\mathrm{cut}} \big) & Z_\mathrm{inj} R_\mathrm{cut} \geq E_\mathrm{inj}
    \end{cases}
\end{split}
\end{equation}
Here, $J_0$ is a normalization constant of the cosmic-ray flux, $a(A_\mathrm{inj})$ denotes the injected fraction of the respective element with mass $A_\mathrm{inj}$, defined below the cutoff. 

\begin{figure}
\centering
\resizebox{0.49\textwidth}{!}{\includegraphics{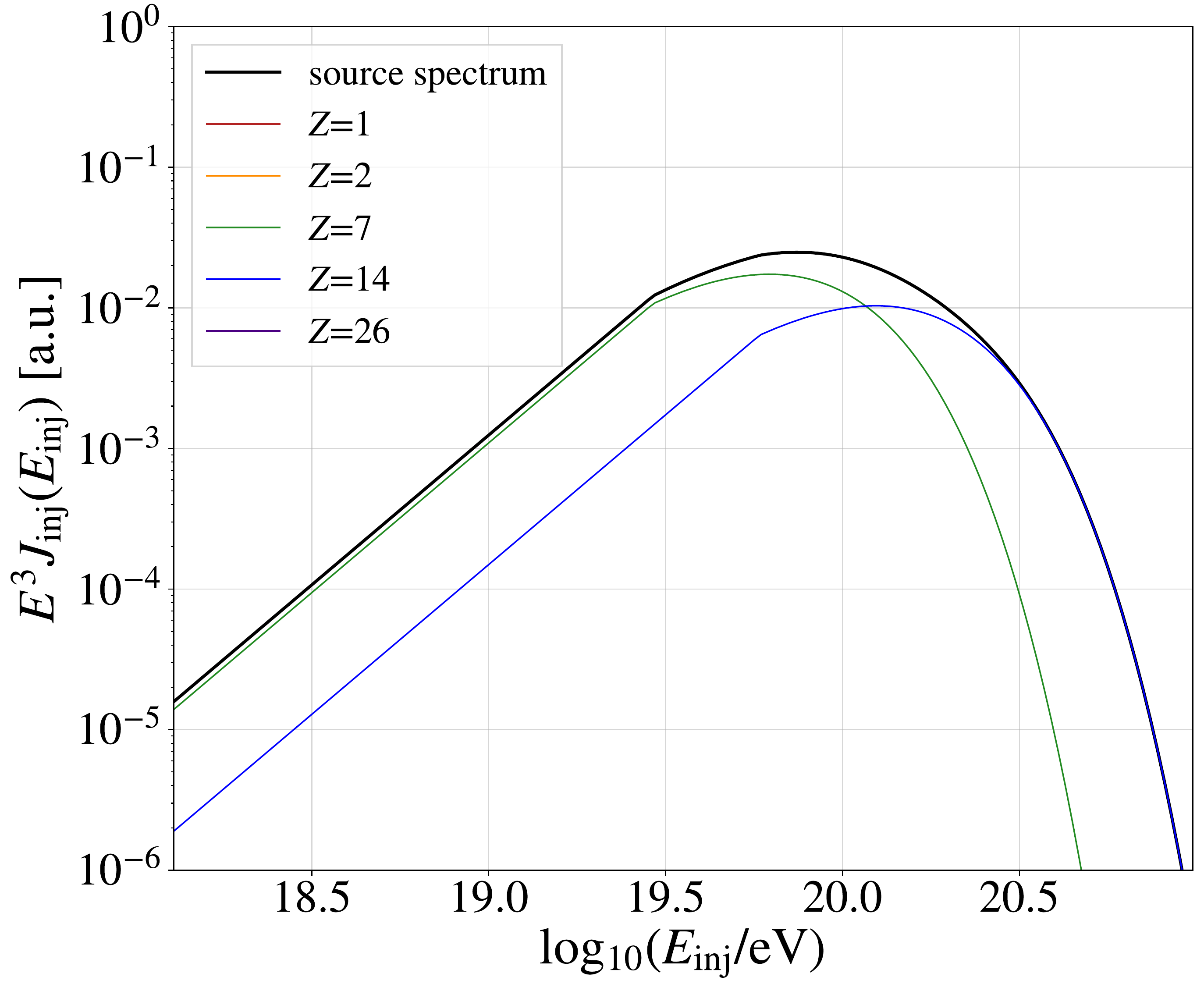}}
\caption{Injected spectrum using the best-fit source parameters from \cite{AugerCF}, following a power law with a broken exponential cutoff above a maximum rigidity as given by eq.~\ref{eq:emission}.}
\label{img:energy_injected}
\end{figure}

Fig.~\ref{img:energy_injected} shows an example of an injected spectrum at the source. Here, the source parameters have been adjusted to best-fit values identified by~\cite{AugerCF}. They are given by the spectral index $\gamma=0.87$, rigidity cutoff $R_\mathrm{cut}=10^{18.62}$~V, fractions of nitrogen $a(\mathrm{N})=88\%$, and silicon $a(\mathrm{Si})=12\%$ following Tables 8 and 9 in~\cite{AugerCF} for a CRPropa3-based model, abbreviated with \texttt{CTG}. This set of parameters will hereafter be referred to as the \textit{benchmark parameters}.

After the cosmic rays have been emitted at the source following the injected spectrum, they propagate through the universe, undergoing interactions before being detected at Earth. The mapping between the observables (the properties of the detected cosmic rays at Earth) and the source parameters at injection ($\gamma$, $R_\mathrm{cut}$, $a(\mathrm{H})$, $a(\mathrm{He})$, $a(\mathrm{N})$, $a(\mathrm{Si})$ and $a(\mathrm{Fe})$) is learned by the cINN. Hence, we first create a database to describe this mapping, which is then used for the network training. The same database is also used for the MCMC evaluation to find the set of source parameters that leads to the best agreement between the simulated observables and the measured data.

This simulation database is created in a modular way using one-dimensional CRPropa3~\cite{CRPropa3} simulations. The data\-base contains $10^6$ simulated cosmic rays for each injected energy between $10^{18}$~eV and $10^{21}$~eV in bins of width $10^{0.02}$~eV and each source distance, binned logarithmically between $1$~Mpc and $5,700$~Mpc in 118 bins. The simulations are performed for each representative element (hydrogen, helium, nitrogen, silicon, and iron) at source injection, and all secondary particles produced on the way to Earth are stored. 
For a homogeneous, isotropic, three-dimensional source population, a uniform distribution of the comoving source distances is expected before propagation effects, which we achieve by reweighting the simulated distances appropriately.

Upon detection at Earth, the cosmic rays are binned into energy bins $e$ of width $10^{0.1}$~eV between $10^{18.0}$~eV and $10^{21.0}$~eV and mass bins $A_\mathrm{det} \in \{1\}$, $[2,4]$, $[5,22]$, $[23,38]$ and~$[39,56]$. Neither the mass nor the charge of the arriving cosmic ray can be directly measured by today's cosmic-ray observatories. Therefore, the depth of the shower maximum $X_\mathrm{max}$ is used as an observable instead, as it relates inversely to the cross-section in air which is connected to the primary cosmic-ray mass. From the detected energies and masses in the simulation database, the expected values for $X_\mathrm{max}$ can be calculated using Gumbel distributions $G$~\cite{deDomenico} and the EPOS-LHC hadronic interaction model~\cite{EPOS}, as in~\cite{AugerCF}.

\begin{figure}
\centering
\resizebox{0.49\textwidth}{!}{\includegraphics{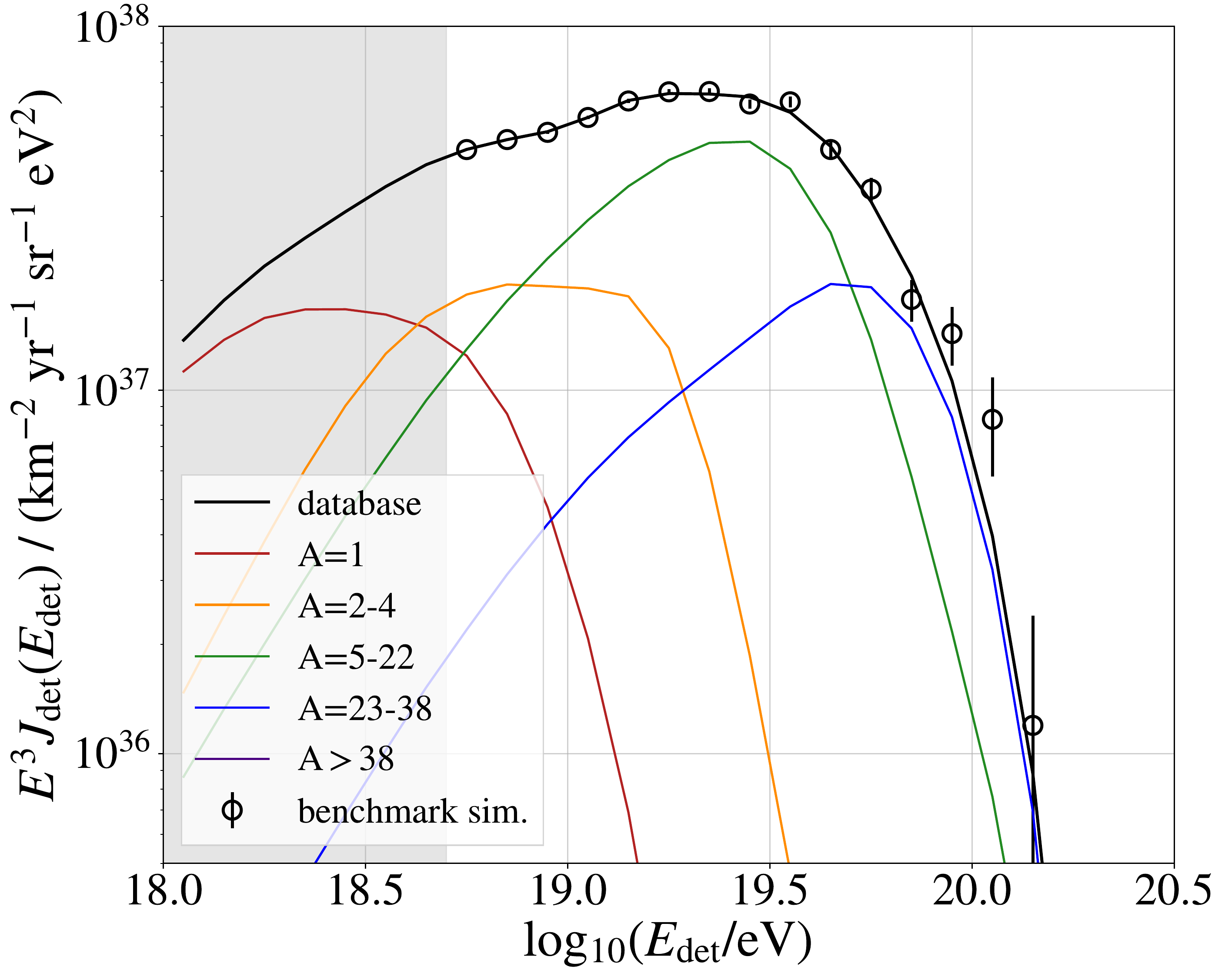}}
\caption{Observed energy spectrum at Earth after injection following Fig.~\ref{img:energy_injected}. As symbols with Poissonian errors, the benchmark simulation spectrum with $\mathcal{O}(70,000)$ events is shown, with different colors for the different detected element groups. The curves depict the prediction by the propagation database scaled to the same number of cosmic rays. The gray area marks the part of the energy spectrum below the threshold at $10^{18.7}$~eV, which is not part of the fit.}
\label{img:energy_detected}
\end{figure}

\begin{figure}
\centering
\resizebox{0.49\textwidth}{!}{\includegraphics{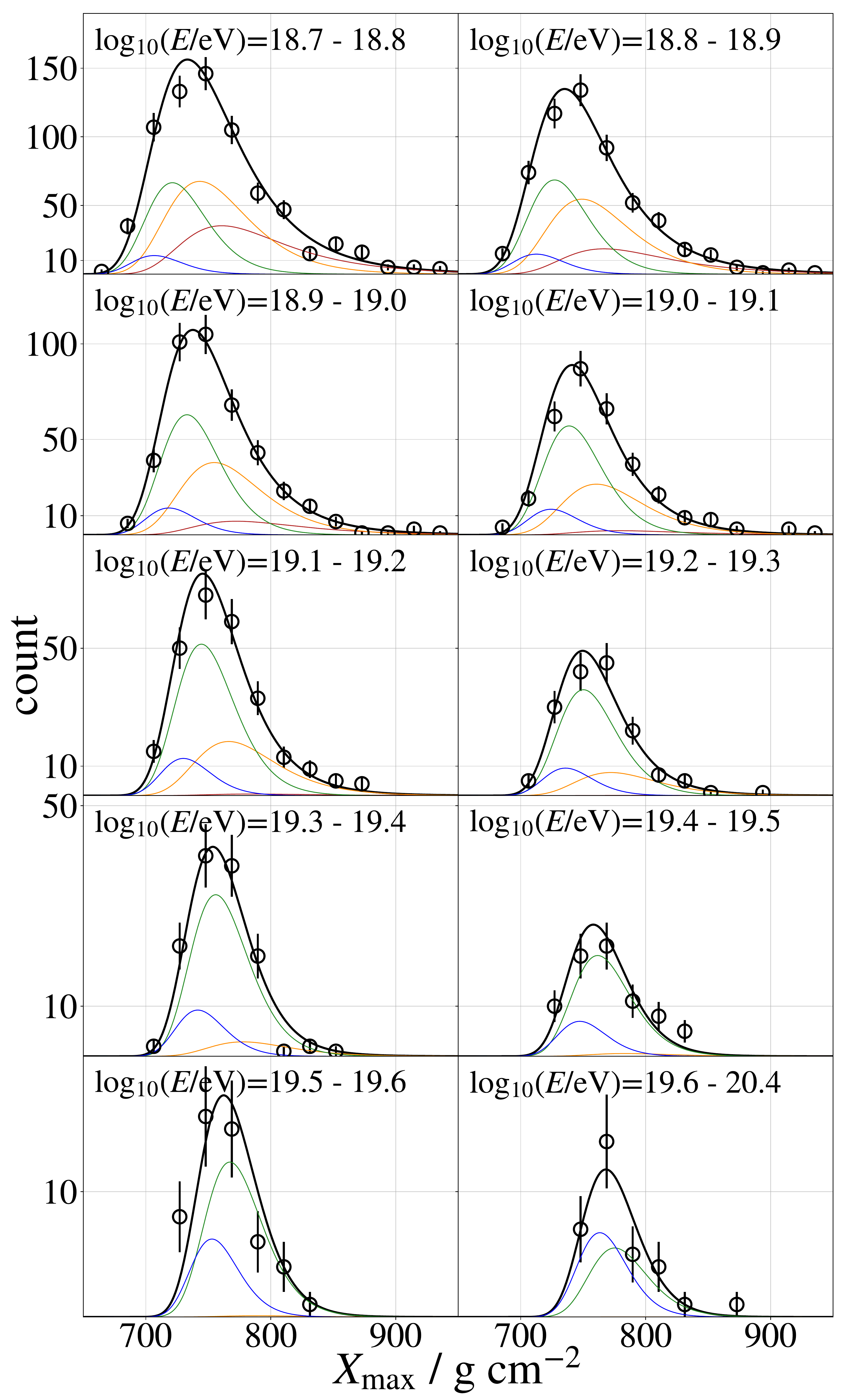}}
\caption{Depth of the shower maximum $X_\mathrm{max}$ distributions in energy bins. The binning is the same as for the energy spectrum with a combined bin above $10^{19.6}$~eV due to the smaller statistics. The benchmark simulation with $\mathcal{O}(2,700)$ events containing an $X_\mathrm{max}$ value is shown as symbols with Poissonian errors. The curves refer to the reweighted distributions from the propagation database, scaled to the same number of cosmic rays. The contributions by the different element groups are color-coded as in Fig.~\ref{img:energy_detected}.}
\label{img:xmax_detected}
\end{figure}

The mapping of the example source spectrum, depicted in Fig.~\ref{img:energy_injected}, to Earth is shown in Fig.~\ref{img:energy_detected} and Fig.~\ref{img:xmax_detected}, where the detected energy spectrum and the detected $X_\mathrm{max}$ distributions can be seen.
One can see that the $X_\mathrm{max}$ histogram is binned two-dimensionally into $X_\mathrm{max}$ bins $x$ between 550~g/cm$^2$ and 1050~g/cm$^2$ of width 20~g/cm$^2$ and similar energy bins $\tilde{e}$ as the observed spectrum with a combined bin above $10^{19.6}$~eV due to the smaller event statistic.
We show not only the shape of the observables predicted by the database for the benchmark parameters as curves, but also one specific simulation from the database with the same number of events as in the data of the Pierre Auger Observatory. For this simulation, we additionally include bin-wise Poisson fluctuations. The spectrum in Fig.~\ref{img:energy_detected} contains $\mathcal{O}(70,000)$ events~\cite{AugerSpectrumICRC2019} and the shower depths histogram in Fig.~\ref{img:xmax_detected} contains $\mathcal{O}(2,700)$ events~\cite{AugerCompositionICRC2019}, both above $10^{18.7}$~eV. This specific simulation will hereafter be referred to as the \textit{benchmark simulation}. 

In both the detected energy spectrum and the shower depth $X_\mathrm{max}$ distributions, the rapid decay of the flux as a function of the energy is visible. It is evident that the propagation has a substantial impact on the cosmic-ray energies, and other elements apart from the injected ones have emerged after interactions and decays. With increasing energy, the composition becomes heavier as expected from the rigidity-dependent acceleration at the source. The shape and location of the $X_\mathrm{max}$ distributions contain information on the composition: lighter cosmic rays can penetrate deeper into the atmosphere as the cross-section for air interactions is smaller, and the shower-to-shower fluctuations are larger than for heavy particles due to the superposition principle~\cite{Kampert2012}.

Altogether, we produced a database for the mapping from the injection at the source to the detected observables at Earth for different source parameters:
\begin{equation}
J_\mathrm{inj}(E_\mathrm{inj}, A_\mathrm{inj}) \xrightarrow{\gamma, R_\mathrm{cut}, a(A_\mathrm{inj})} (E_\mathrm{det}, X_\mathrm{max}^\mathrm{det})
\label{forward_model}
\end{equation}
In the following sections, we will describe how the MCMC and the cINN methods use this database for determining the source parameters. Afterward, both methods will be applied to the \textit{benchmark simulation} in sec.~\ref{sec:mcmc_vs_cinn}.

\section{MCMC method for inference}
\label{sec:MCMC}
Markov Chain Monte Carlo methods are used to determine an unknown posterior probability density function (pdf) by sampling from it. The basis for all MCMC methods is Bayes theorem, which connects the unknown posterior pdf $p(\theta|y)$ of the fit parameters $\theta$ given the data $y$ to the likelihood of the data given the fit parameters $p(y|\theta)$ multiplied by the prior pdf $p(\theta)$:
\begin{equation}
    p(\theta|y) = \frac{p(y|\theta) \ p(\theta)}{p(y)} \ \propto \ p(y|\theta) \ p(\theta)
\end{equation}
Here, $p(y)$, which is often called the \textit{Bayes integral}, is generally hard to calculate. Using Bayes' theorem in MCMC sampling has the advantage that $p(y)$ does not have to be known, as the shape of the posteriors can be determined without this normalization. To this end, one uses only prior knowledge $p(\theta)$ on the source parameters multiplied by the likelihood $p(y|\theta)$. This likelihood corresponds to the \textit{forward} direction described in sec.~\ref{sec:introduction}, so it is often known or can be predicted by models. This does not require any derivatives or integrals to be calculated as is the case, for example, for minimizers~\cite{MCMC}.

In our case, the likelihood can be calculated using the propagation database described in section~\ref{sec:astrophysical_model}, which predicts the energy spectrum and $X_\mathrm{max}$ distributions (corresponding to $y$) from the source parameters (corresponding to $\theta$).
Specifically, we use the same likelihood function $\mathcal{L} = \mathcal{L}_E \cdot \mathcal{L}_{X_\mathrm{max}}$ as in~\cite{AugerCF}. It contains a Poissonian likelihood $\mathcal{L}_E$ for the energy spectrum, which compares the predicted spectrum calculated from the simulation database (event counts $p$ in the energy bin $e$) to the benchmark simulation (corresponding event counts $k$):
\begin{equation}
    \mathcal{L}_E = \prod_e ~\frac{(p^e)^{k^e}}{k^e!} \ \exp{(-p^e)}
\end{equation}
The information on the energy spectrum is already used in the energy likelihood, so a multinomial likelihood $\mathcal{L}_{X_\mathrm{max}}$ is used for the $X_\mathrm{max}$ distributions:
\begin{equation}
\label{eq:likelihood_xmax}
    \mathcal{L}_{X_\mathrm{max}} = \prod_{\tilde{e}} k^{\tilde{e}}! \prod_x ~\frac{(G^{\tilde{e},x})^{k^{\tilde{e},x}}}{k^{\tilde{e},x}!}
\end{equation}
Here, $k^\mathrm{\tilde{e}, x}$ again describes the measured number of events in each energy bin $\tilde{e}$ and $X_\mathrm{max}$ bin $x$, and $G^\mathrm{\tilde{e}, x}$ represents the Gumbel distributions for the respective bin as in~\cite{AugerCF}.

For the fit of the benchmark simulation, we let the sampler run 50,000 discarded burn-in steps followed by 50,000 sampling steps in 15 different chains using the Metropolis-Hastings algorithm \cite{Metropolis,Hastings}. Convergence is ensured by calculating the Gelman-Rubin coefficient $R~$\cite{GR}. We use the same flat bounded prior distributions for the source parameters as in~\cite{AugerCF}. Each chain runs for around 4-6 hours on a CPU.

\section{Conditional invertible neural networks (cINN)}

An alternative to MCMC sampling is a new method that uses deep learning techniques, introduced in~\cite{INN} as an \textit{invertible neural network} (INN) and extended to the \textit{conditional} setup in~\cite{BayesFlow} and~\cite{ImageGeneration}. The idea is based on the concept of normalizing flows, by which an invertible mapping is created between the physics parameters $\theta$, the source parameters in our case, and internal network parameters, referred to as latent variables $z$. The remarkable property of this bijective mapping is that it preserves probability. 

\subsection{Introduction to cINNs}

\begin{figure*}
\centering
\resizebox{0.6\textwidth}{!}{\includegraphics{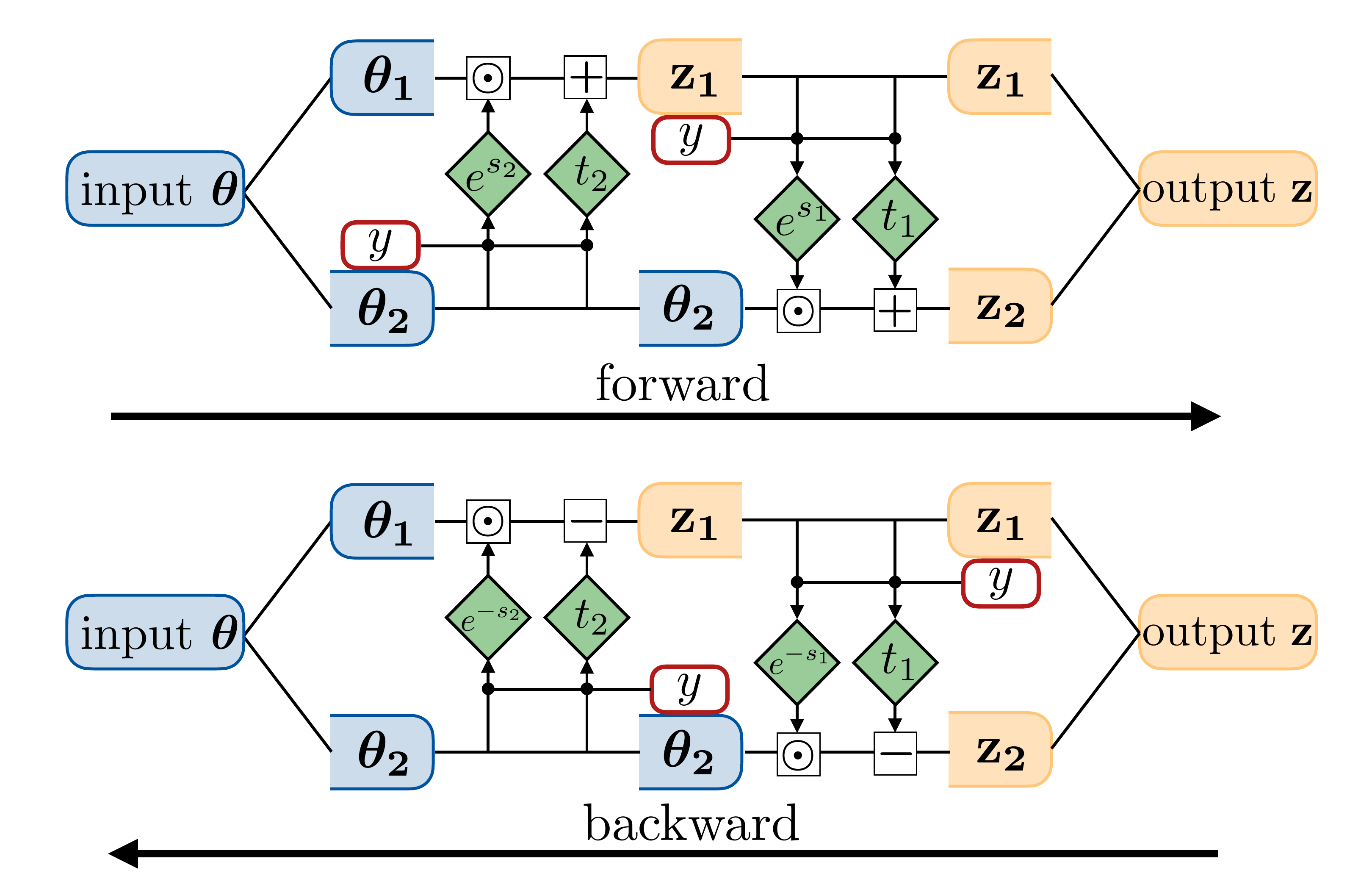}}
\caption{Structure of the reversible block used for the conditional invertible neural network. It can be evaluated in two directions. The upper part shows the training mode or forward direction, the lower part displays the evaluation mode or backward direction.}
\label{img:cINN}
\end{figure*}

To create the invertible mapping $z=f(\theta)$ between the internal network parameters $z$ and the physics parameters $\theta$, \textit{reversible blocks}~\cite{RNVP} are used. Fig.~\ref{img:cINN} shows a schematic sketch. It is based on the architecture introduced in \cite{RNVP} and \cite{GLOW} and can be evaluated in both the forward as well as the backward direction. In the forward direction, the input vector $\mathbf{\theta}$ is first split into two halves. The output vector $z = [z_1, z_2]$, where $z_1$, $z_2$ correspond to the first and second half of the output vector, is determined as follows:
\begin{equation}
    \begin{aligned}
        z_1 &= \theta_1 \odot \exp (s_2(\theta_2)) + t_2(\theta_2) \\
        z_2 &= \theta_2 \odot \exp (s_1(z_1)) + t_1(z_1)
    \end{aligned}
\label{eq:reversibleblock_forward}
\end{equation}
where $\odot$ refers to element-wise multiplication and the mappings $s_i()$ and $t_i()$ can be arbitrarily complicated and do not have to be invertible themselves. The mappings $s_i, t_i$ are in general represented by additional neural networks. In the GLOW~\cite{GLOW} setup, the mappings $s_i, t_i$ are computed by a single subnetwork for each $i$. 
The inverse of the affine transformation can be easily obtained since the exponential function prevents division by zero and the subnetworks are always evaluated in the same direction:
\begin{equation}
    \begin{aligned}
        \theta_2 &= (z_2 - t_1(z_1)) \odot \exp (-s_1(z_1)) \\
        \theta_1 &= (z_1 - t_2(\theta_2)) \odot \exp (-s_2(\theta_2))
    \end{aligned}
\end{equation}
In the conditional setup the observables $y$, the energy spectrum, and the depth of shower maximum distributions are fed into the reversible block as conditional inputs. This concept is shown in Fig.~\ref{img:cINN}. The mappings $s_i(...)$ and $t_i(...)$ then become $s_i(..., y)$ and $t_i(..., y)$. Before the transformation with the subnetworks $s_i, t_i$, the observables are concatenated to the input of the subnetworks. These blocks are then used to construct the  \textit{conditional invertible neural network}.
\\
For the training, we also need a suitable loss function, which was introduced in~\cite{BayesFlow}. The goal is to train a network that represents a mapping of a distribution in the latent space $p(z)$ to the true posterior space $p(\theta|y)$ (backward direction). Thus, we want to minimize the difference between the cINN posterior $p_\phi(\theta|y)$, where $\phi$ denote the network parameters, and the true posterior $p(\theta|y)$. The Kullback-Leibler divergence $\mathbb{KL}$ provides a measure on the difference of two probability distributions and is used as the basis of the loss $L$:
\begin{equation}
    \begin{aligned}
        L &=  \mathbb{KL} \big(p(\theta | y) \;\| \;p_\phi(\theta | y)
        \big) \\
        &= \mathbb{E}_{\theta \sim p(\theta | y)} \big(\log p(\theta | y)-\log p_\phi(\theta | y)\big) \\
        &= \mathrm{const.} + \mathbb{E}_{\theta \sim p(\theta |y)}\big(-\log p_\phi(\theta | y)\big) 
    \end{aligned}
\end{equation}
Here, $\mathbb{E}$ denotes the expectation value, with parameter values $\theta$ sampled from the distribution $p(\theta | y)$.
In the last step, the true posterior distribution is constant with respect to the network parameters and can thus be omitted in the loss function. Next, we apply the concept of probability conservation $p_\phi(\theta|y) \mathrm{d}\theta = p(z) \mathrm{d}z$ to transform the network posterior to the latent space:
\begin{equation}
    \begin{aligned}
    L &= \mathbb{E}_{\theta \sim p(\theta | y)} \big( -\log p_\phi(\theta | y)\big) \\
    &= \mathbb{E}_{\theta \sim p(\theta | y)} \bigl(- \log \big( p(z) \cdot | \det \left( \frac{\partial \,z}{\partial \theta} \right) | \big) \bigr) \\
    &= \mathbb{E}_{\theta \sim p(\theta | y)} \bigl(- \log \big( p(z) \big) - \log\big(| \det \left( \frac{\partial \,z}{\partial \theta} \right) | \big) \bigr)
    \end{aligned}\label{eq:loss}
\end{equation}
The Jacobian $\partial z/\partial \theta$ of the reversible blocks (Fig.~\ref{img:cINN}), which map from the physics parameter space $\theta$ to the latent space $z$ via $z = f(\theta)$, turns out to be a triangular matrix. This simplifies the calculation of the determinant substantially. 
To see the argument, we decompose the transformation of the reversible block in eq.~(\ref{eq:reversibleblock_forward}) into two functions $f_1$ and $f_2$.
Using as an example $f_1$
\begin{equation}
    f_1(\theta) =
    \begin{cases}
    z_1 = \theta_1 \odot \exp(s_2(\theta_2)) + t_2(\theta_2) \\
    \theta_2 = \theta_2\; ,
    \end{cases}
\end{equation}
its Jacobian is calculated as follows:
\begin{align}
    \det \frac{\partial f_1(\theta)}{\partial \theta} &=
    \det
    \begin{pmatrix}
        \frac{\partial z_1}{\partial \theta_1} & \frac{\partial z_1}{\partial \theta_2} \\
        \frac{\partial \theta_2}{\partial \theta_1} & \frac{\partial \theta_2}{\partial \theta_2}
    \end{pmatrix} \\
    &=
    \det
    \begin{pmatrix}
        \mathrm{diag} ( \exp \left(s_2({\theta}_2) \right) ) & \frac{\partial {z}_1}{\partial {\theta}_2} \\
        0 & \mathbb{I}
    \end{pmatrix} \\
    &= \prod_j \exp \left(s_{2, j}({\theta}_2) \right)
\end{align}
Equivalently, the Jacobian of $f_2(\theta)$ is calculated, resulting in the total determinant:
\begin{equation}
    \begin{aligned}
        | \det \left( \frac{\partial \,z}{\partial \theta} \right) |
        &= | \frac{\partial f_1(\theta)}{\partial \theta} \frac{\partial f_2(\theta)}{\partial \theta} | \\
        &= \prod_j \exp(s_{2, j}(\theta_2)) \cdot \exp(s_{1, j}({z}_1)) \\
        &= \exp \big({\sum_j s_{2, j}(\theta_2)+ s_{1, j}({z}_1)}  \big)
    \end{aligned}
\end{equation}
Now one can decide on the form of the distribution $p(z)$ that is enforced on the latent variables. To simplify the loss function as in~\cite{BayesFlow}, we choose a unit Gaussian distribution, denoted in one dimension by $p(z) = p(f(\theta)) = \exp (- f(\theta)^2/2)$. With the logarithmic functions in eq. (\ref{eq:loss}), this results in the following loss function for all parameter dimensions and the two subnetworks, averaged over $m$ training datasets:
\begin{equation}
    L = \frac{1}{m} \sum_{i=1}^m \bigl( \frac{1}{2} \Vert f(\theta_i) \Vert^2 - \sum_{l=1}^2 \sum_{j} s_{l, j} \bigr)
\end{equation}

\subsection{cINN for inference}

The framework we use for the cINN for the inference of source parameters using the energy spectrum and the depth of shower maximum distribution as observables is called Framework for Easily Invertible Architectures (FrEIA)~\cite{ImageGeneration} and is based on the PyTorch library~\cite{pytorch}. The network consists of six reversible blocks with a GLOW~\cite{GLOW} subnetwork structure. The mappings $s_{1, 2}, t_{1, 2}$ are represented by three fully connected layer transformations with an internal width of 256 with ReLU activation functions. Like in \cite{ImageGeneration} and \cite{StellarParameters}, prior to the exponential transformation of $s_i$, a non-linear transformation according to $\tilde{s}=0.636\, \alpha\, \arctan (s/ \alpha)$ is applied to support stable training, here using $\alpha=1.9$. After each reversible block, a permutation layer is used to enhance mixing between the different latent variables. The conditions $y$ are the binned energy and shower maximum values (see Fig.~\ref{img:cINN}). 

The training data are created with the aforementioned data\-base for mapping the source parameters $\theta$, namely the spectral index $\gamma$, the maximum rigidity $R_\mathrm{cut}$, and the five elemental fractions $a(\mathrm{H})$, $a(\mathrm{He})$, $a(\mathrm{N})$, $a(\mathrm{Si})$ and $a(\mathrm{Fe})$, to the detected observables on Earth. The spectral index and the maximum rigidity have already been constrained by~\cite{AugerCF}, which we use to limit our training data to reasonable pairs of $(\gamma, R_\mathrm{cut})$ around the found minimum. The elemental fractions can be sampled uniformly using $(5-1)=4$ representative variables~\cite{simplex} to satisfy the condition that the sum equals one. 

$1,000,000$ training samples and $100,000$ validation samples with their corresponding energy spectrum and depth of shower maximum are generated. An interesting note is that a factor of $10$ fewer training examples compromised the results. Before entering these into the network, they have to be preprocessed.
The spectral parameters $\gamma$ and $\log_{10}(R_\mathrm{cut}/\, \mathrm{V})$ are transformed to values between 0 and 1. The elemental fractions, being physically constrained to non-negative values, are converted with the inverse logistic function before the training.
For the energy spectrum, we use the aforementioned $17$ energy bins $e$ (Fig.~\ref{img:energy_detected}). During the training, each of the energy bin contents is modified according to a Poisson distribution with the number of events corresponding to the typical event statistics measured by the Pierre Auger Observatory (see section~\ref{sec:astrophysical_model}). This is important for the network to learn to evaluate different scenarios with the underlying statistical fluctuations.
Afterward, the bin content of each energy bin is multiplied by $E^3$. This helps flatten the steeply decreasing spectrum measured at Earth (Fig.~\ref{img:energy_detected}), and its effectiveness in improving the reconstruction quality of the cosmic-ray source parameters was checked. The network is given this modified bin content of the 17 energy bins as conditional input, where the sum over all bins is normalized to one.

The depth of shower maximum distributions are binned into the bins $\tilde{e}, x$ as described in sec.~\ref{sec:astrophysical_model}. Again, as for the energy spectrum, each bin content is altered using a Poisson distribution with reduced statistics, as expected by the measurements at the Pierre Auger Observatory. Here, by normalizing the $X_\mathrm{max}$ distribution in each energy bin to unity we remove the energy spectrum information that is already used as a separate observable. To feed this (10 x 24) matrix into the network, we use flattening which results in a one-dimensional array with $240$ entries.

The 17 energy bins are entered into the first 3 layers of the network as conditional input, and the $X_\mathrm{max}$ distributions into the last three layers. We verified that the information of both observables is indeed used by the network. The training of the network takes thirty hours on a GPU and the evaluation of a single scenario can be completed in seconds.

\section{Determination of source parameters with the cINN compared with the MCMC method}
\label{sec:mcmc_vs_cinn}

In the following, we evaluate the \textit{benchmark simulation} presented in sec.~\ref{sec:astrophysical_model} using the MCMC and the cINN methods. Both methods yield posterior distributions of the fit parameters, which can be used to determine the most probable value and the uncertainty on the parameters by the $68 \, \%$ interval as well as unveil correlations between the parameters. 

Even though both methods can be used to characterize the posteriors, they use inherently different mathematical bases for it. The MCMC uses a likelihood, which is engineered according to the experimental statistics of the observables measured at Earth, as presented in sec.~\ref{sec:MCMC}. The difference between the predicted and measured observables is minimized when maximizing the likelihood, and Bayes' theorem ensures convergence of the sampled space to the posteriors~\cite{MCMC}. 

The cINN, on the other hand, uses a likelihood-free inference where a loss function minimizes the distance between the true posterior distributions of the source parameters and the posterior distribution as predicted by the network. It does not ensure agreement of the network prediction with the measured observables at Earth, which is only implicitly achieved by the agreement of the source parameter posteriors.

\begin{figure}
\centering
\resizebox{0.45\textwidth}{!}{\includegraphics{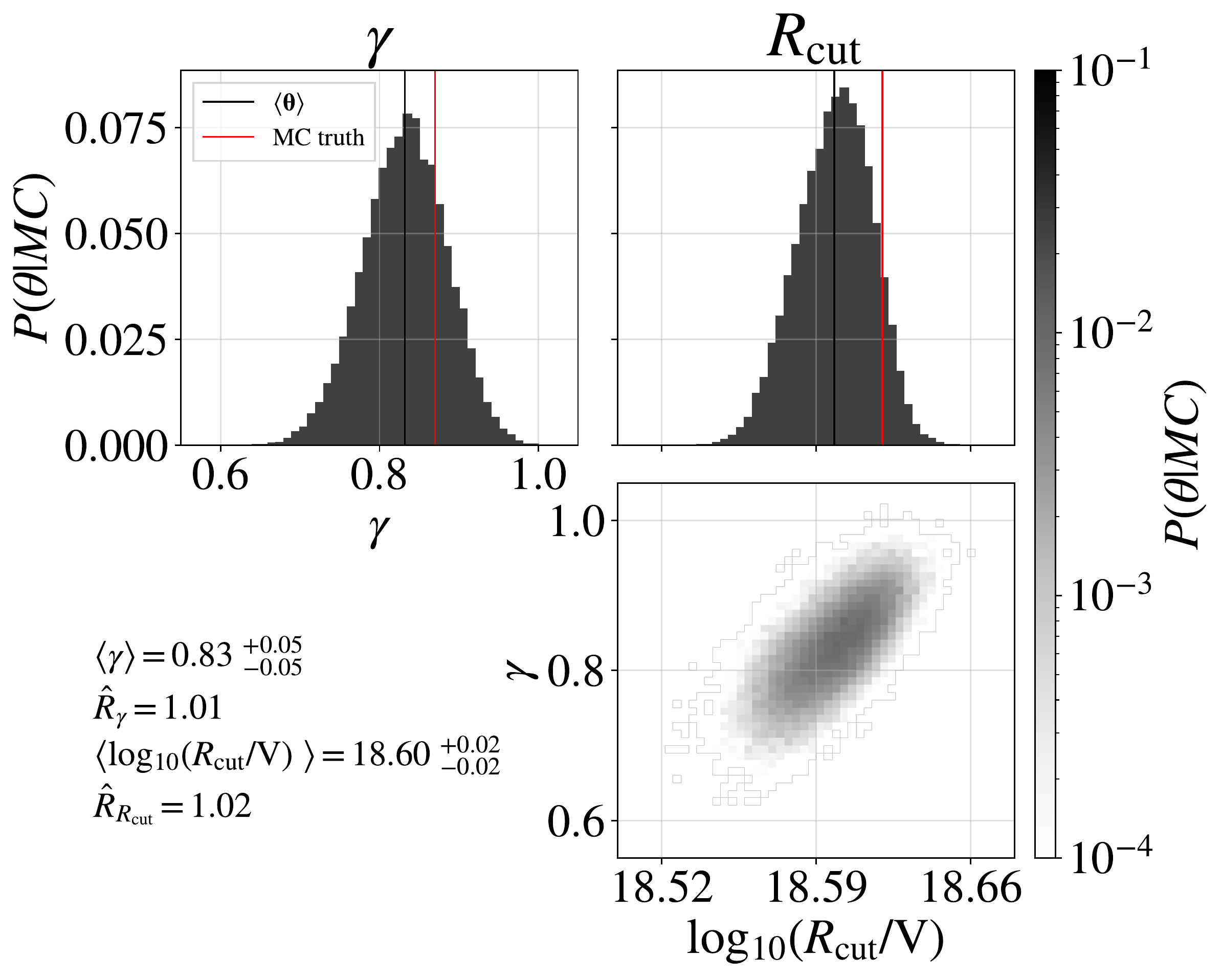}}
\resizebox{0.45\textwidth}{!}{\includegraphics{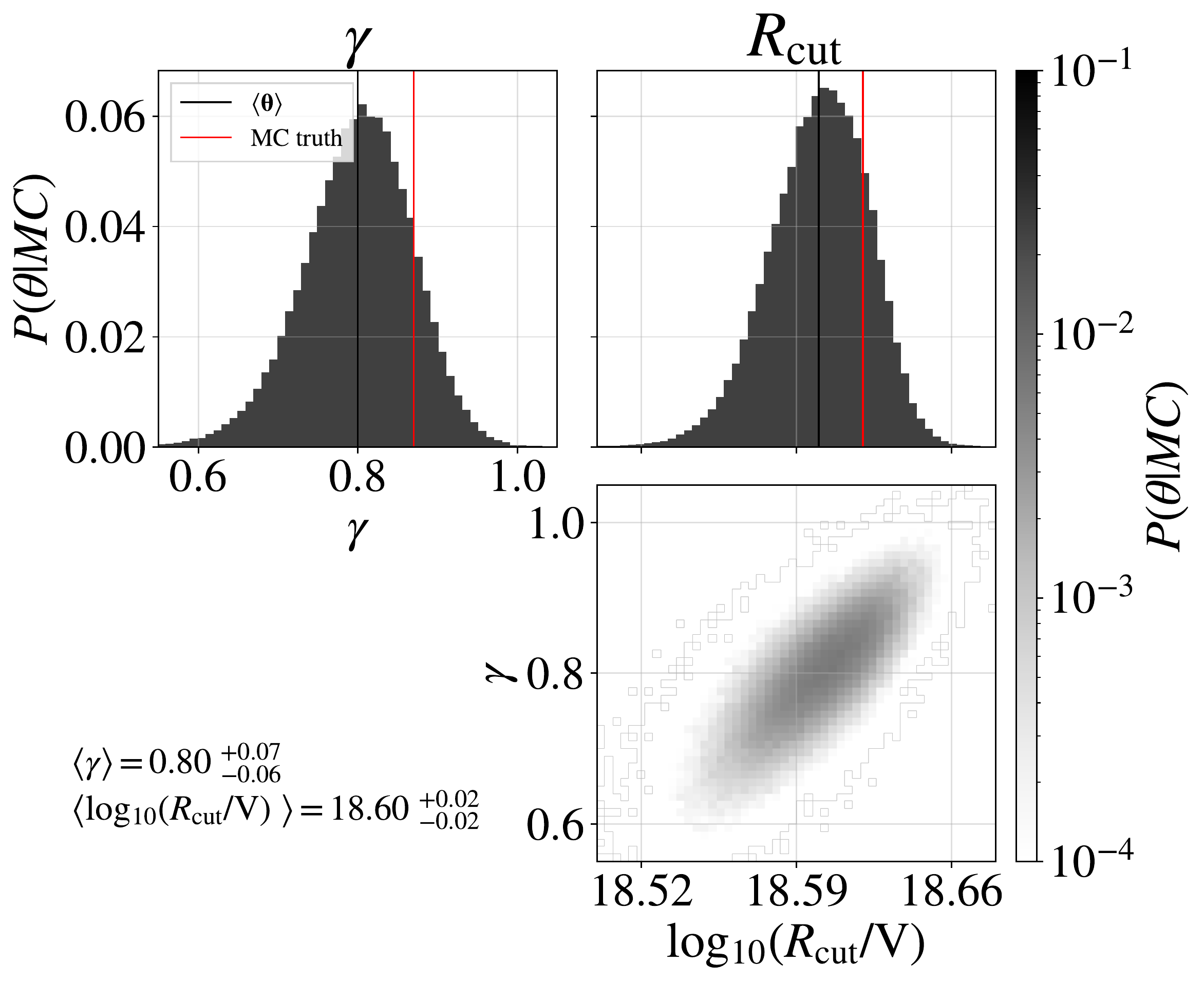}}
\caption{Posterior distributions obtained with the MCMC \textit{(upper)} and the cINN \textit{(lower)} methods of the spectral index $\gamma$ and the rigidity cutoff $\log_{10}(R_\mathrm{cut}/\, \mathrm{V})$. The mean of the distributions is shown by the black solid curve and the \textit{true} underlying value of the \textit{benchmark simulation} is marked by the red curve. In the lower left of the plots of both methods, the mean and standard deviation of both parameters are shown, and for the MCMC also the Gelman-Rubin index $R$.}
\label{img:posterior}
\end{figure}

Fig.~\ref{img:posterior} shows the posterior distributions for the spectral index $\gamma$ and the rigidity cutoff $R_\mathrm{cut}$ from the MCMC in the upper part and from the cINN in the lower part, respectively. The posterior distributions of $\gamma$ and $\log_{10}(R_\mathrm{cut}/\, \mathrm{V})$ are generally similar for both methods: the one-dimensional histograms show a symmetric distribution with the \textit{true} benchmark parameters within one standard deviation of the posterior mean. Both methods find a positive correlation between the parameters, shown in the two-dimensional lower histograms. One can see that both methods slightly underestimate the two parameters. It was checked that this finding depends on the specific scenario chosen; hence, using different Poissonian variations of the observables leads to slightly shifted posteriors for both methods.

One can see that the widths of the posteriors are slightly larger for the cINN than for the MCMC. For the cINN the widths of the posteriors depend on the size of the training dataset and the training time of the network, therefore it must be ensured that the network is trained with a sufficiently large training dataset for a long enough time. The same applies for the MCMC, where a sufficient number of chains have to be run with enough sampling steps. This can be ensured by keeping the Gelman-Rubin coefficient $R$~\cite{GR} close to one, which is shown in the figures. Also, we trained multiple networks with different initializations and compared the posteriors, which appear to be quite similar. The example shown here is obtained from the cINN with the lowest validation loss value.

\begin{figure}
\centering
\resizebox{0.45\textwidth}{!}{\includegraphics{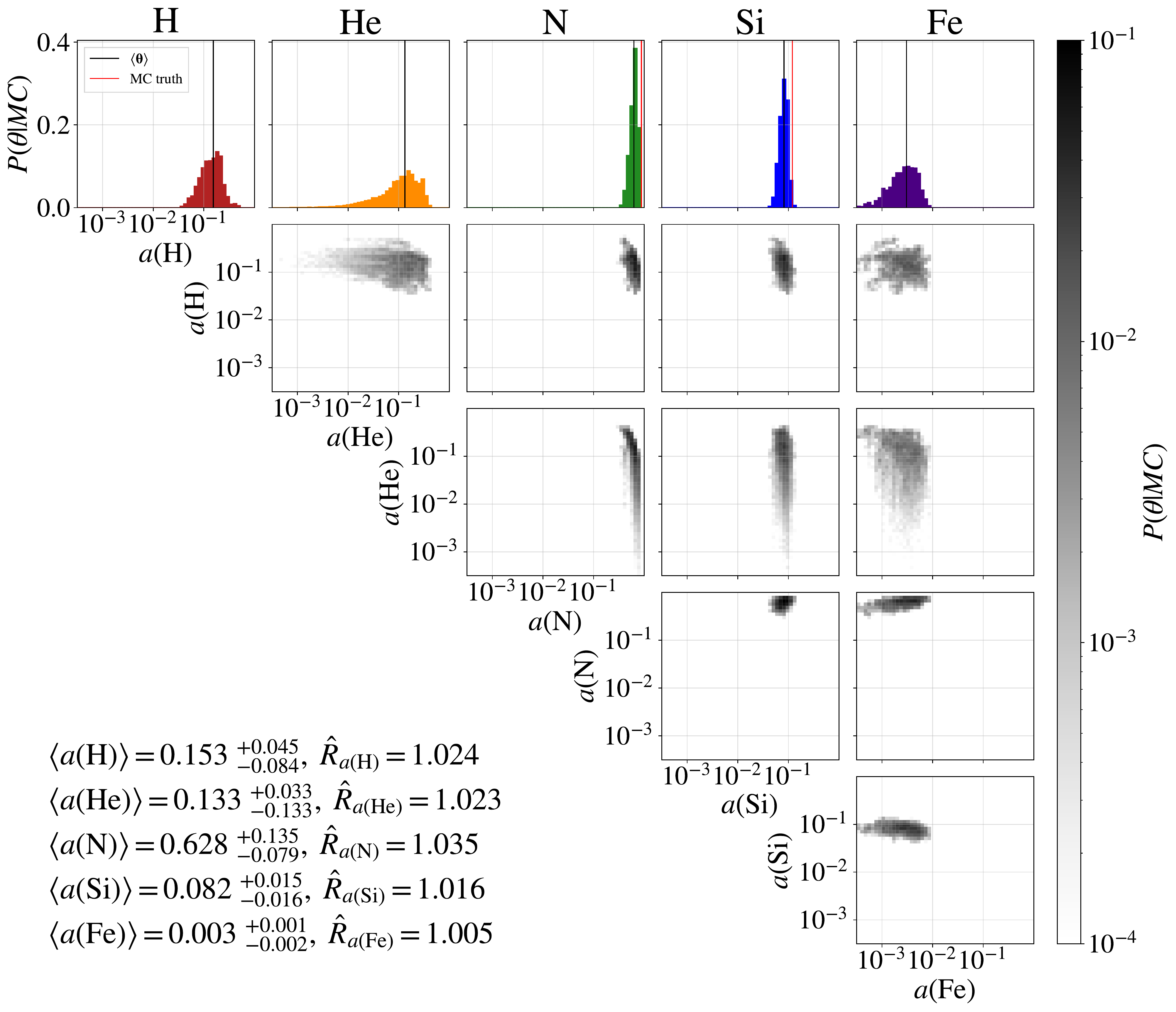}}
\resizebox{0.45\textwidth}{!}{\includegraphics{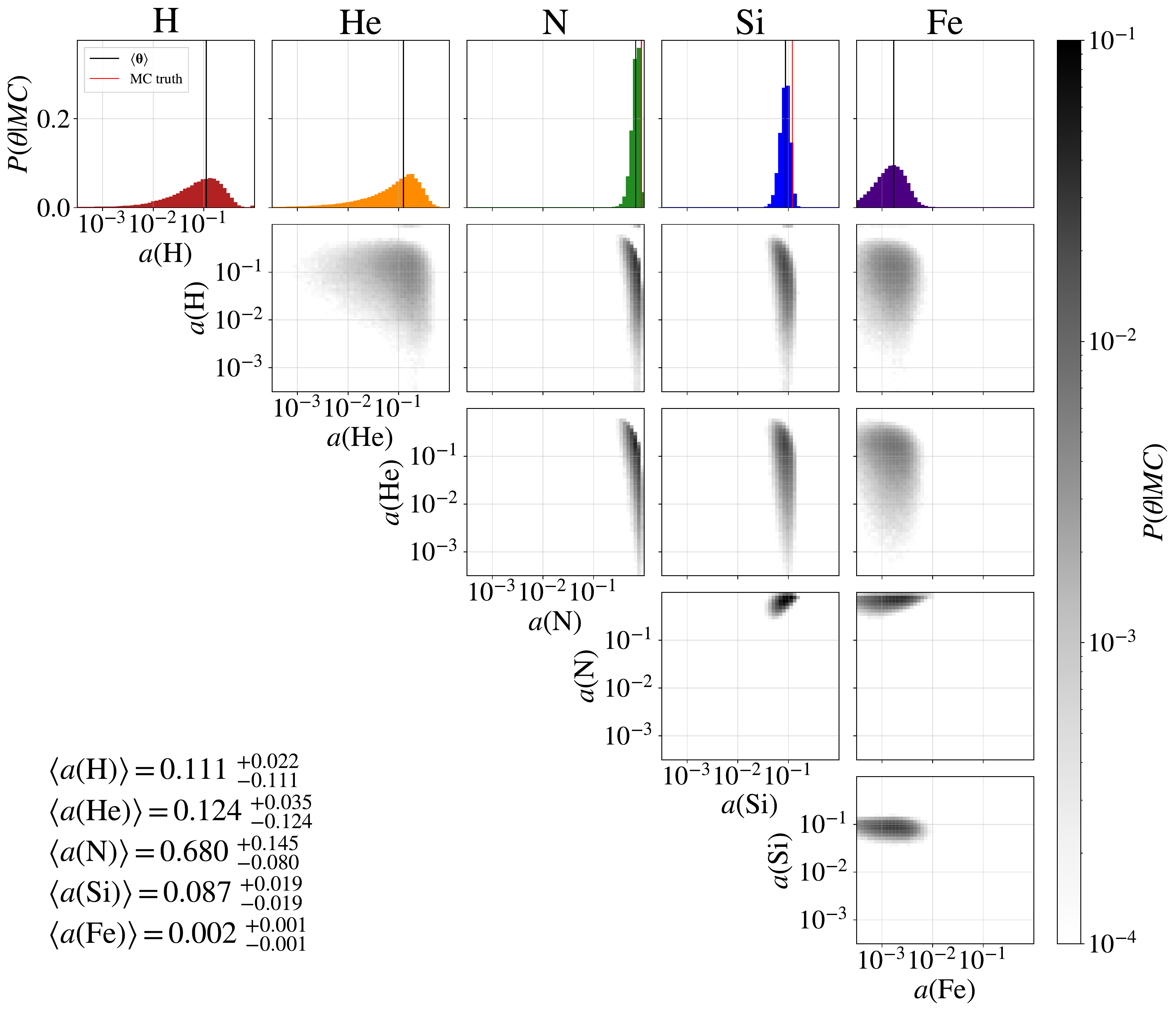}}
\caption{Posterior distributions of the composition fractions obtained with the MCMC \textit{(upper)} and the cINN \textit{(lower)} methods. The mean of the distributions is shown by the black solid curve and the \textit{true} underlying value of the \textit{benchmark simulation} is marked by the red curve. In the lower left of the plots of both methods, the mean and standard deviation of both parameters are shown and for the MCMC also the Gelman-Rubin index $R$.
}
\label{img:posteriors_fractions}
\end{figure}

Fig.~\ref{img:posteriors_fractions} shows the posterior distributions for the composition fractions of the five representative elements. One can see that both methods again lead to similar results in general. Both are able to identify that the composition at the source is dominated by nitrogen and silicon and that the iron fraction is tiny. The posteriors of the lighter elements are very broad, ranging down to zero contribution. This indicates that those parameters are more difficult to determine, which is attributable to the fact that the cutoff energy (cf. Fig.~\ref{img:energy_injected}) for hydrogen is at $E_\mathrm{cut} = Q_e \cdot R_\mathrm{cut} = 10^{18.62}$~eV ($Q_e$ denotes the elementary charge) and the one for helium is at $2 Q_e \cdot R_\mathrm{cut} = 10^{18.92}$~eV, which means that almost no light primaries are expected to survive above the energy threshold of the observables at $10^{18.7}$~eV.

The correlations between the fractions look similar using both methods. The histograms for the MCMC look less smooth than for the cINN; this is due to the fact that we have several chains combined into one posterior. The Gelman-Rubin coefficient $R$ is close to one for all parameters, indicating a convergence of the chains, but the posteriors could still become slightly smoother with more sampling steps or more chains. 

\begin{figure}
\centering
\resizebox{0.45\textwidth}{!}{\includegraphics{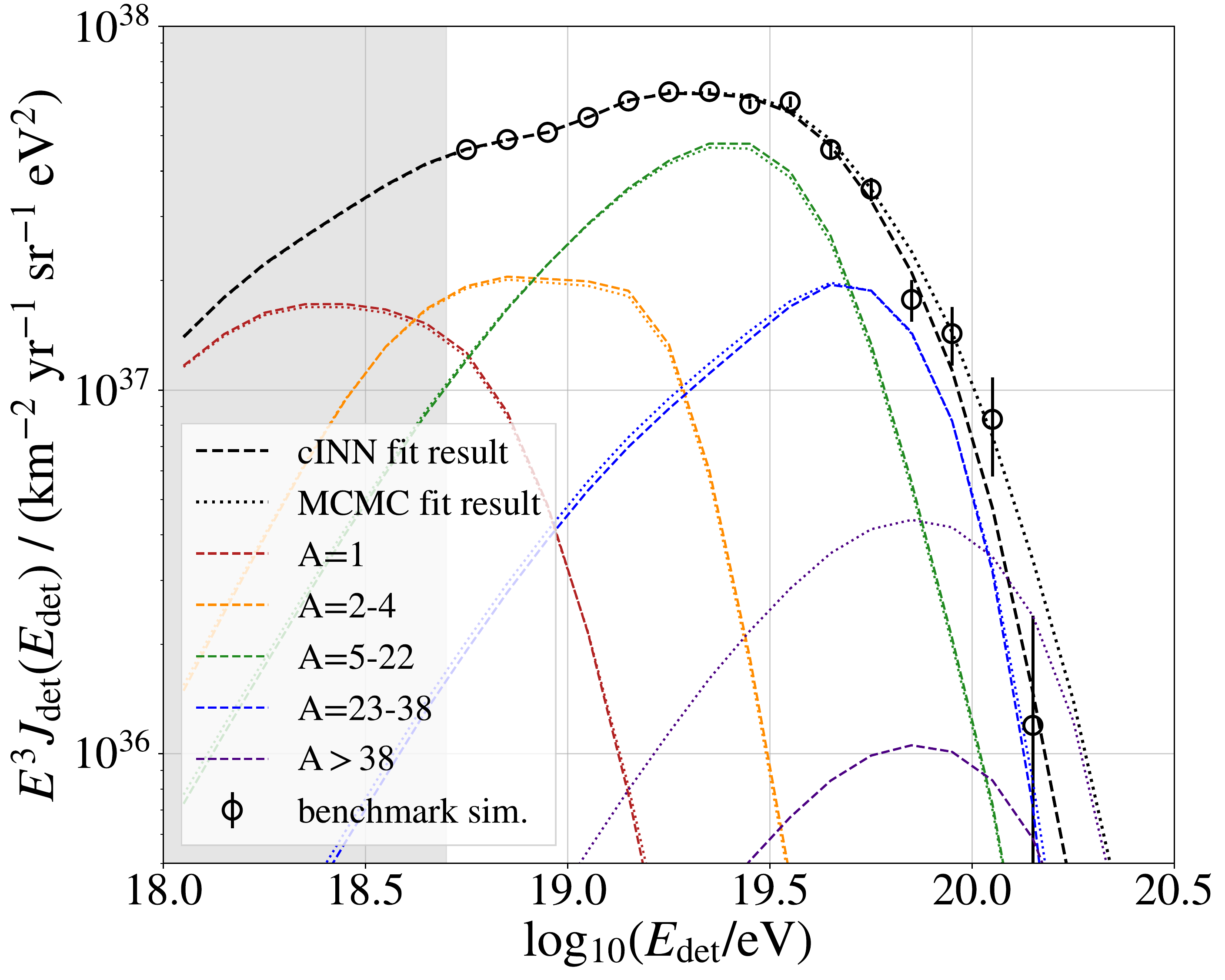}}
\caption{Modeled energy spectra for the cINN (dashed) and the MCMC (dotted) using the predicted source parameters as given in Fig.~\ref{img:posterior} and \ref{img:posteriors_fractions}. Both methods are able to find a set of source parameters that describe the benchmark simulation energy spectrum, depicted as black symbols including Poissonian error bars. Also, the individual element contributions predicted at Earth, shown in different colors for different mass groups, agree with the ones of the benchmark simulation in Fig.~\ref{img:energy_detected}.}
\label{img:reconstruction}
\end{figure}

In general, we observe very similar posterior distributions generated by the two different methods. One can additionally compare the reconstructed observables as shown in Fig.~\ref{img:reconstruction} exemplarily for the energy spectrum. In comparison with Fig.~\ref{img:energy_detected}, one can see that both methods yield good agreement between the modeled spectrum and the benchmark simulation. The same applies to the $X_\mathrm{max}$ histograms (not shown here). The level of agreement can be quantified by calculating the deviance $D$~\cite{AugerCF}, which is two times the negative log-likelihood ratio of the model, and the \textit{saturated} model that would describe the data perfectly. For the two observables, the energy spectrum and the $X_\mathrm{max}$ distributions, we use the likelihood functions which are used directly for the MCMC sampling, as given in sec.~\ref{sec:MCMC}. For the MCMC we achieve a deviance of $D^\mathrm{MCMC} = D_E + D_{X_\mathrm{max}} = 14.8 + 123.6 = 138.4$ and for the cINN $D^\mathrm{cINN} = 15.5 + 125.6 = 141.1$. We obtain reasonable, quite similar values for both methods, indicating a good description of the observables.

\subsection{Stability of the cINN results}

\begin{figure}
\centering
\resizebox{0.24\textwidth}{!}{\includegraphics{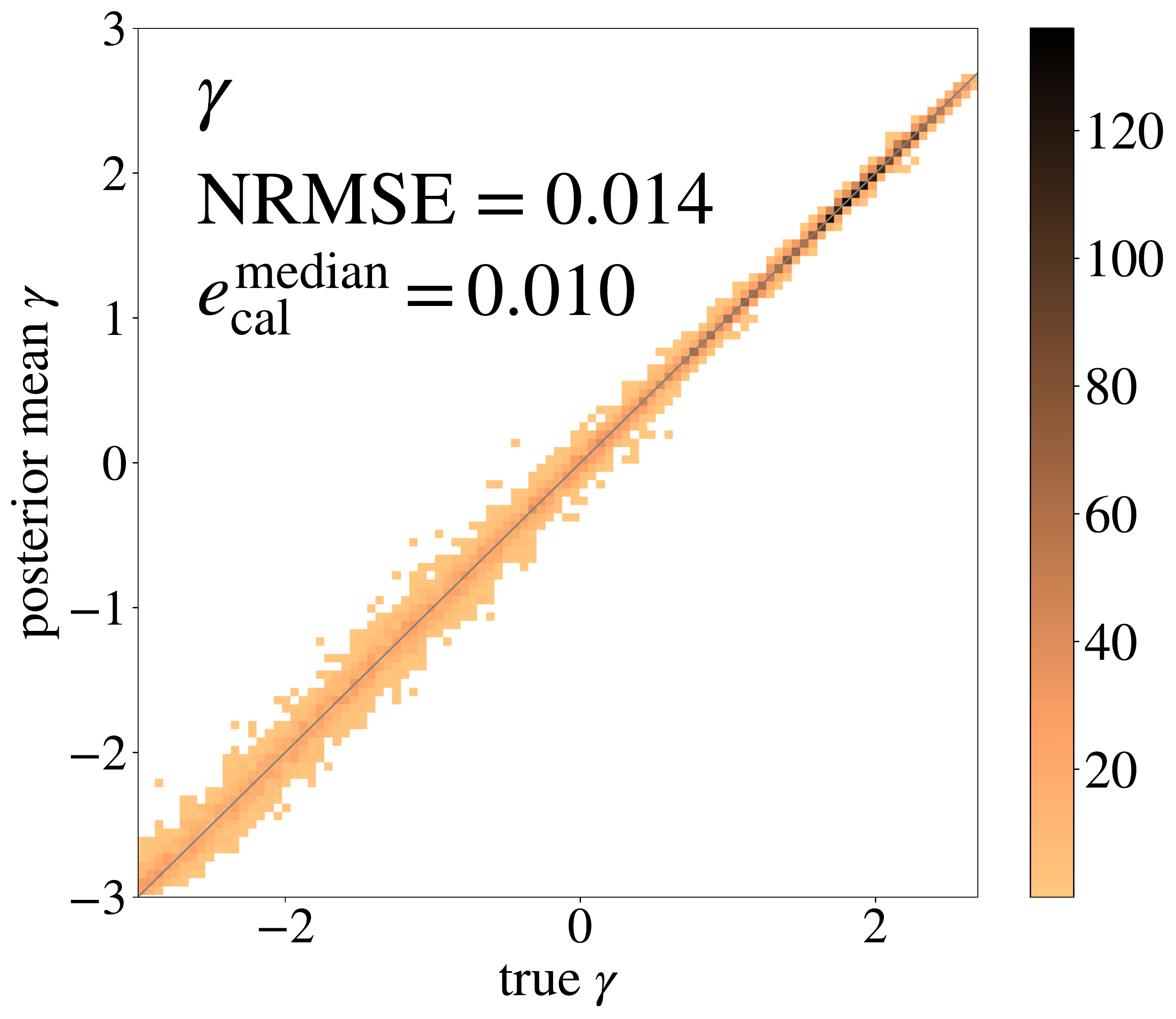}}
\resizebox{0.24\textwidth}{!}{\includegraphics{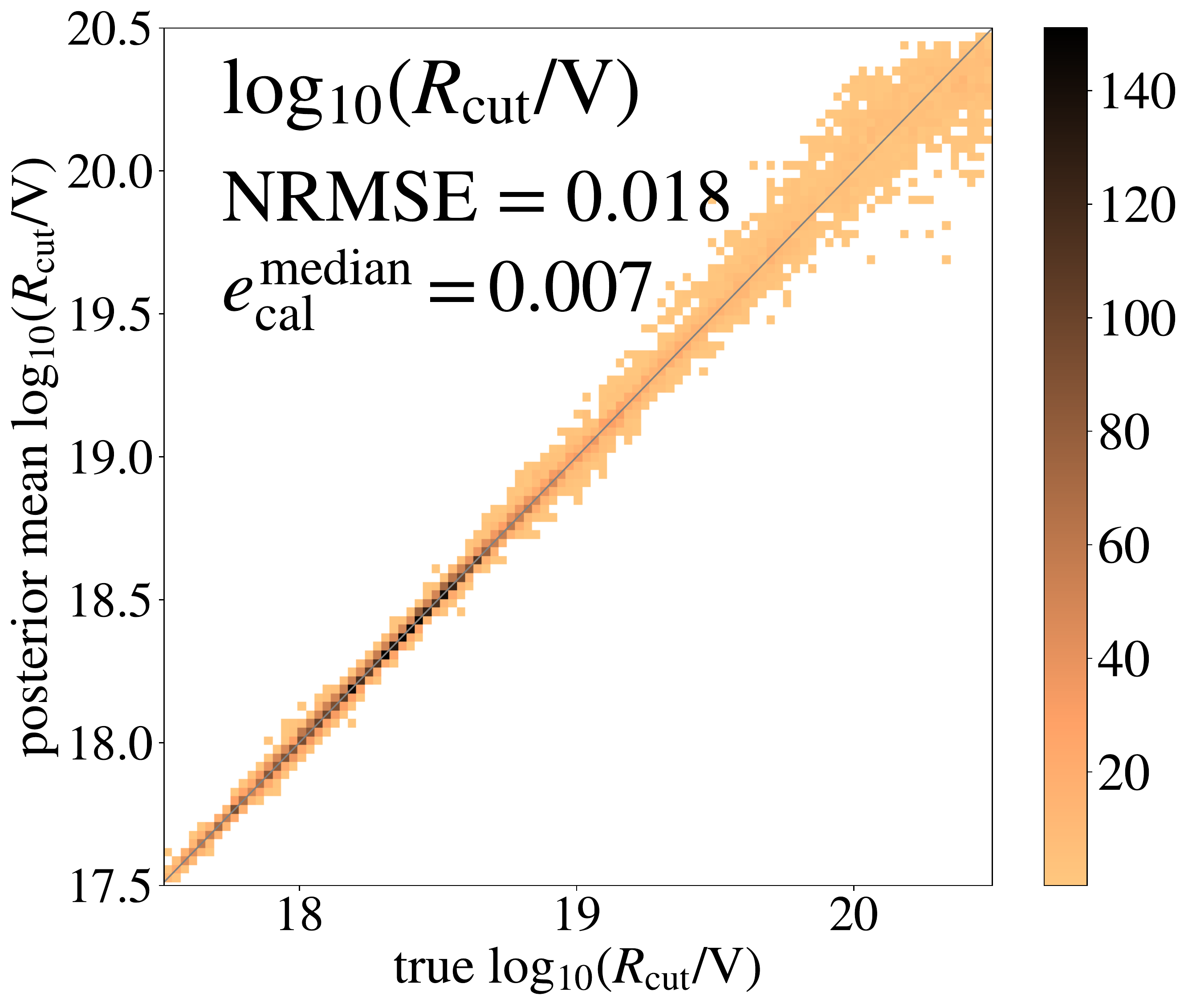}}
\resizebox{0.24\textwidth}{!}{\includegraphics{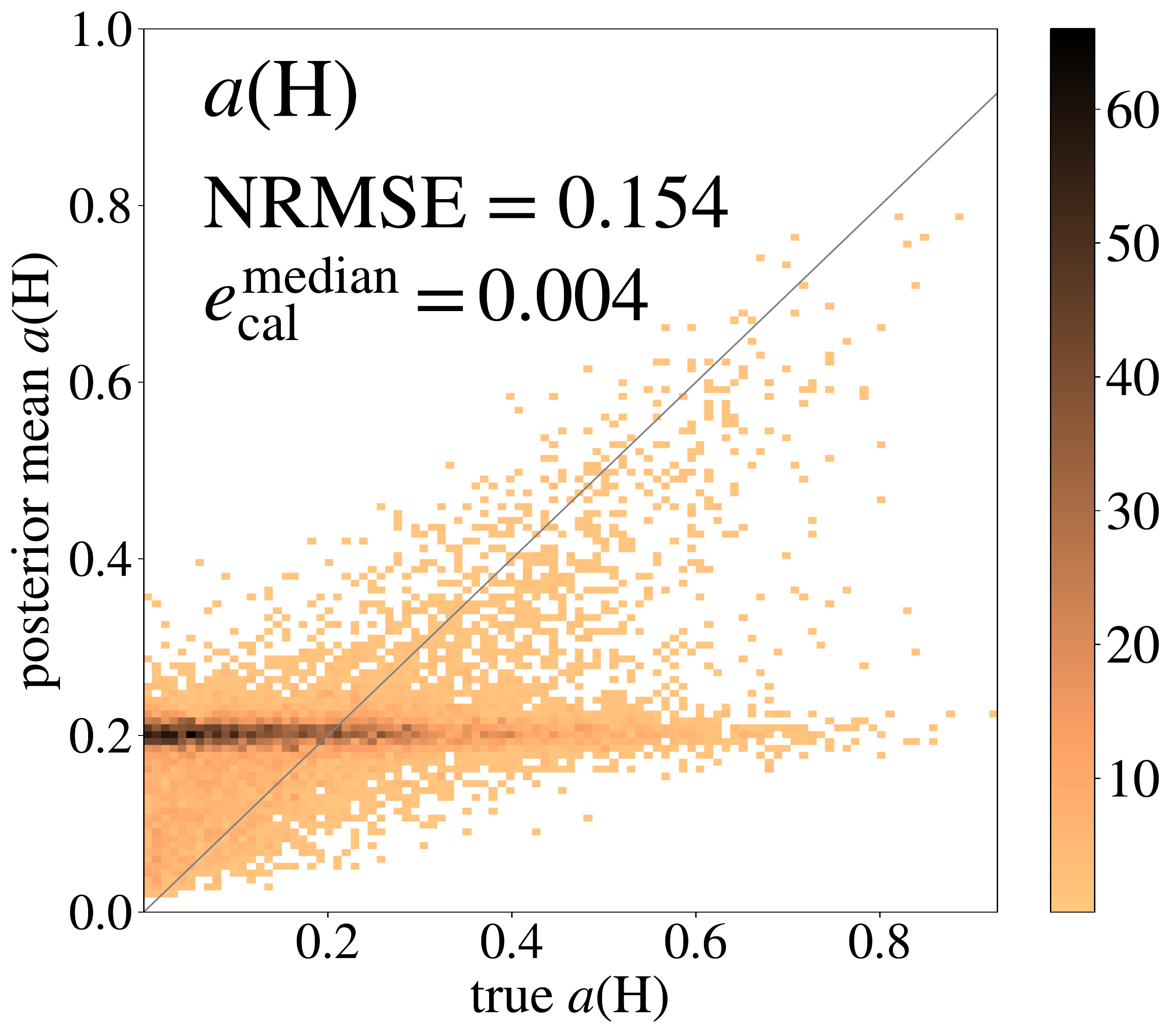}}
\resizebox{0.24\textwidth}{!}{\includegraphics{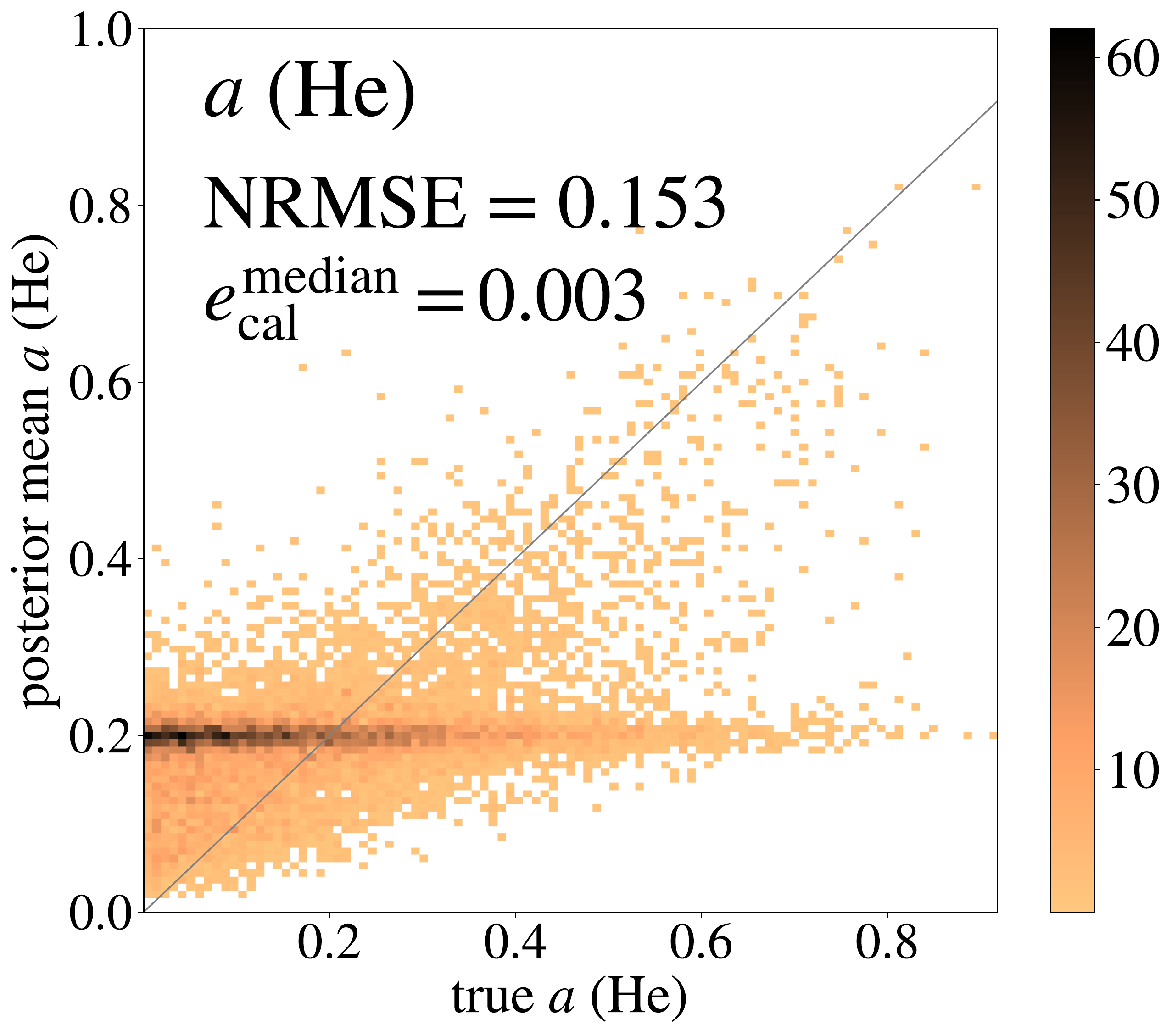}}
\resizebox{0.24\textwidth}{!}{\includegraphics{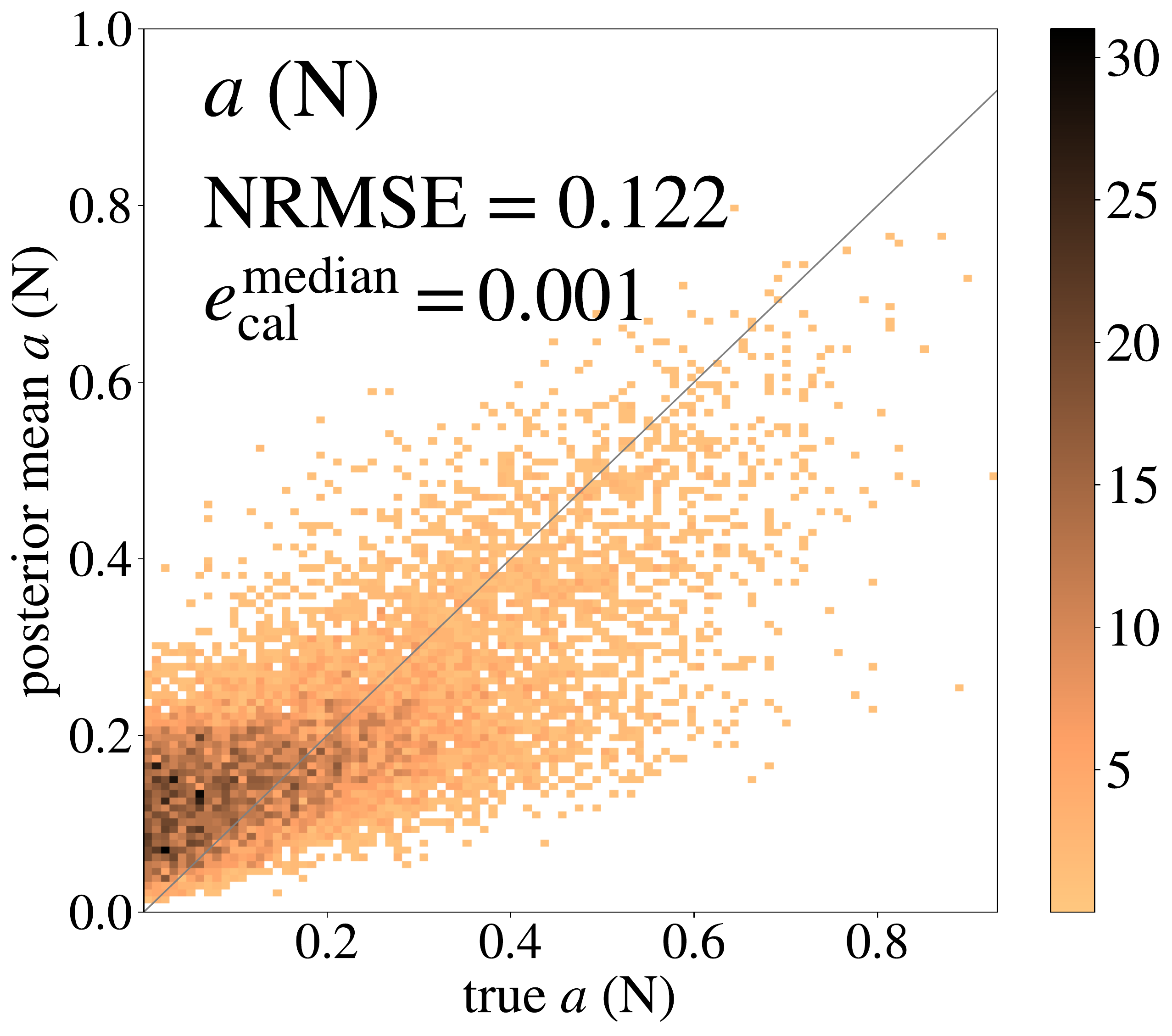}}
\resizebox{0.24\textwidth}{!}{\includegraphics{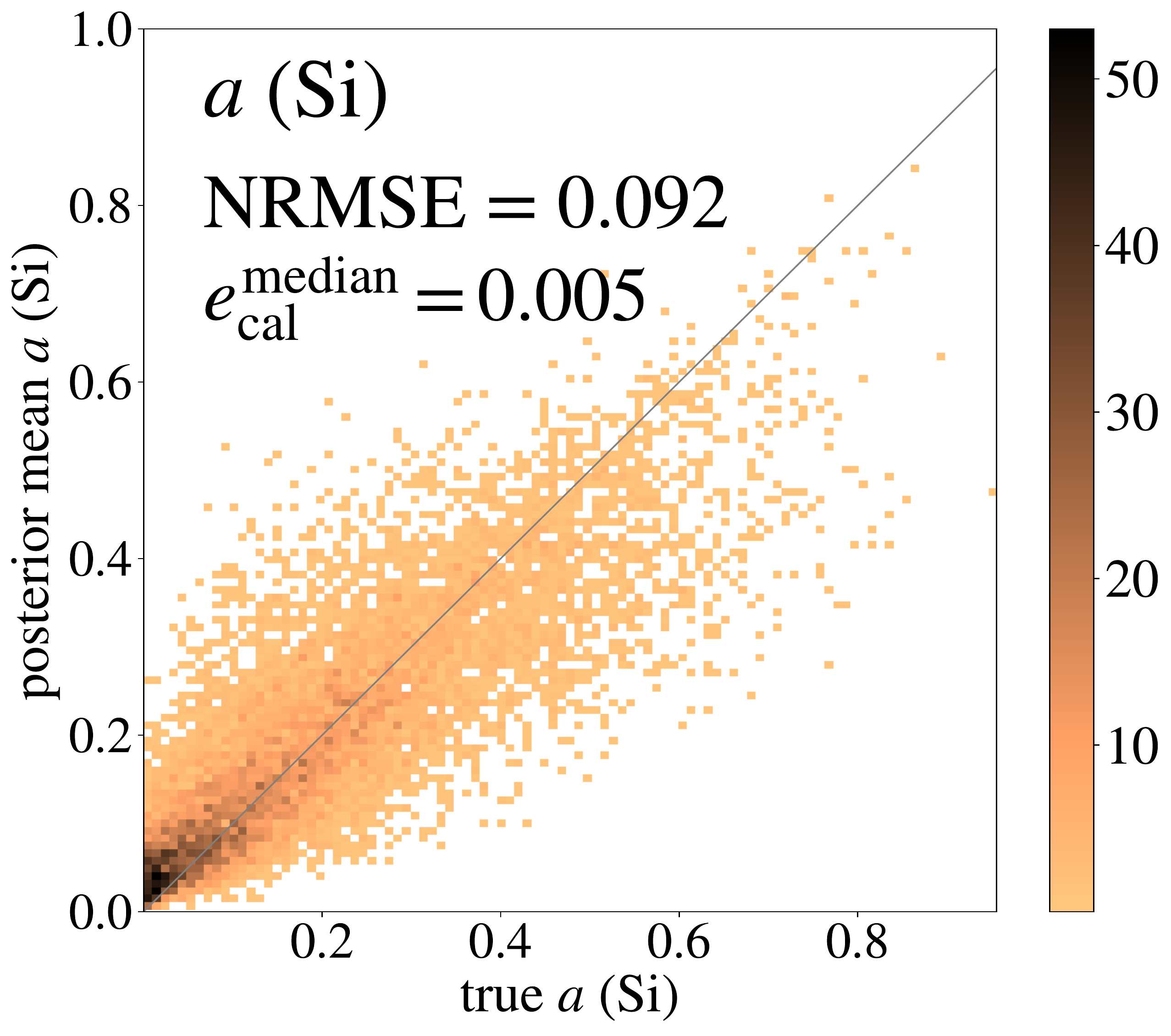}}
\resizebox{0.24\textwidth}{!}{\includegraphics{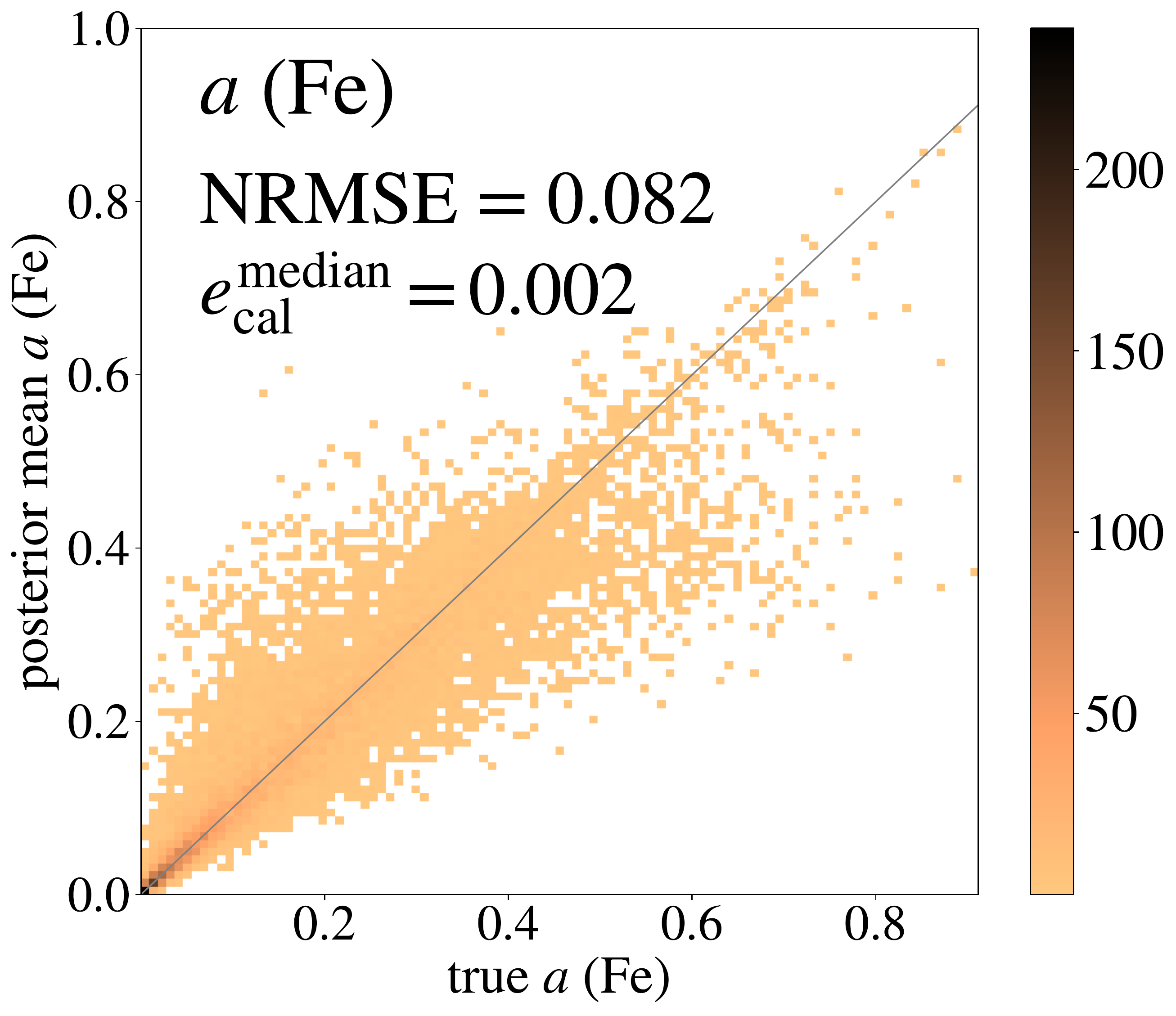}}
\caption{$2D$ histograms of the $7$ source parameters $\gamma$, $R_\mathrm{cut}$ and $a_{(\mathrm{H, He, N, Si, Fe)}}$ using $10^4$ test datasets. The true value of the simulation is shown on the horizontal axis, the mean of the cINN posterior on the vertical axis. The gray straight line represents the position of perfect agreement.}
\label{img:uncertainty}
\end{figure}

To evaluate the performance of the new cINN method in more detail, we evaluated not only this single simulation, but also a test dataset of $10,000$ simulations. This extensive test was performed only with the cINN and not with the MCMC, as the computing time for the MCMC exceeds reasonable times within our computational resources. Fig.~\ref{img:uncertainty} shows two-dimensional histograms of the source parameters. The true value of each simulation in the test set is shown on the $x$-axis, the mean of the cINN posterior is indicated on the $y$-axis.

For the spectral index $\gamma$ and the cutoff rigidity $R_\mathrm{cut}$ we see good agreement between the mean estimate and the true simulation value which can be confirmed by the small normalized root mean square error
\begin{equation}
    \mathrm{NRMSE} = \cfrac{\sqrt{\cfrac{\sum_{i=1}^N(\langle \theta_i \rangle - \theta_i)^2}{N}}}{\max (\theta)-\min (\theta)} \, ,
\end{equation}
which is $0.014$ for the spectral index $\gamma$ and $0.018$ for the rigidity cutoff. No far outliers are found and the small widening of the distribution for larger values of $R_\mathrm{cut}$ is due to the degeneracy of high rigidity values for larger spectral indices $\gamma$ as revealed by the previous data analysis presented in~\cite{AugerCF}.

For the composition fractions, the lighter element fractions cannot be reconstructed, as described above. In this case, the cINN mostly just predicts the average value for five elements $1/5$.
The NRMSE value is larger than $0.15$ for the light elements. For the heavier elements, the reconstruction ability improves and the NRMSE decreases to $0.082$ for iron.

Additionally, the widths of the posterior distributions can be examined. For this, we calculate the median calibration error $e_\mathrm{cal}$ following~\cite{StellarParameters}. The calibration error is defined as the difference between a confidence level $q$ and the actual fraction of observations $q_\mathrm{inliers}=N_\mathrm{inliers}/N$ of the whole test dataset of size $N$ within that $q$-confidence interval. We calculate the calibration error for a range of confidence intervals $(0.01, \, 0.99)$ in $0.01$ steps and take the median over the absolute values. Appropriate posterior distributions would result in values close to zero. We reach a median calibration error of $0.001$ to $0.010$ for all parameters as given in Fig.~\ref{img:uncertainty}, confirming suitable widths of the posterior distributions from the cINN for the whole test dataset. This also applies to the light element fractions with often too large posterior means, indicating that the cINN predicts a suitably large uncertainty for these unrecoverable parameters, as was also seen in Fig.~\ref{img:posteriors_fractions}.

\section{Conclusion}

We presented the application of a new method using deep learning techniques, the so-called conditional Invertible Neural Network (cINN), to a scenario from astroparticle physics constraining characteristic cosmic-ray source parameters. Using the energy spectrum and shower depth distributions on Earth as observables, the network is able to assess posterior distributions of the source parameter space. These allow not only a best-fit value to be estimated, but also uncertainties, possible degeneracies, and correlations between the parameters to be unveiled. The accuracy of the approach has been tested and verified to provide promising results for a large phase space of the source parameters. Given the speed of the method, it is easily possible to extend the scenario to more observables and more characterizing parameters of cosmic-ray sources. This allows for potential future applications of the technique. 

Additionally, we compared the method with the conventional MCMC method on a specifically simulated scenario similar to the measurements of the Pierre Auger Observatory. The two inference methods use rather different techniques. While the cINN method aims at matching the true and the predicted distributions of the source parameters, the MCMC method is based on a likelihood analysis where the simulations are adapted to the observed data distributions. Nevertheless, we found good reconstruction of the source parameters within one standard deviation for both methods and an overall agreement of the posterior distributions. 

Training of the cINN takes approximately thirty hours while the evaluation of several test scenarios can be done instantaneously. For the MCMC however, each chain runs for around five hours, and several chains are needed to ensure convergence. Each new test scenario has to be evaluated single-handedly with the MCMC, making the cINN significantly more computationally effective overall.

\section*{Acknowledgments}

This work is supported by the Ministry of Innovation, Science, and Research of the State of North Rhine-Westphalia, and by the Federal Ministry of Education and Research (BMBF).

\newcommand{\doi}[1]{\href{https://doi.org/#1}{\url{#1}}}
\def\etal{et~al.\xspace}

\def\etal{et~al.\xspace}
\newcommand{\journal}[5]{\href{https://doi.org/#5}{\textit{#1}\ \textbf{#2}\ (#3)\ #4}}
\newcommand{\journalpytorch}[5]{\href{#5}{\textit{#1}\ \textbf{#2}\ (#3)\ #4}}
\newcommand{\journalieee}[3]{\href{#3}{\textit{#1}\ {#2}}}
\newcommand{\arXiv}[1]{\href{https://arxiv.org/abs/#1}{arXiv:\nolinkurl{#1}}}

\end{document}